# Observation of a Large Photo-response in a Single Nanowire (Diameter ~30 nm) of Charge Transfer Complex Cu:TCNQ

Rabaya Basori[†], Kaustuv Das[†], T. Phanindra. Sai[†], Prashant Kumar[‡], K.S.Narayan[‡], Arup K. Raychaudhuri[†]*

[†]Theme Unit of Excellence in Nano Device Technology, S.N. Bose National Centre for Basic Sciences, Salt Lake, Kolkata-700098, India

E-mail: (arup@bose.res.in)

[‡]Jawaharlal Nehru Center for Advanced Scientific Research, Jakkur Bangalore-560 064, India

KEYWORDS: Cu:TCNQ, Single Nanowire, Photo-response, MSM device.

ABSTRACT: We report for the first time large photoresponse in a single NW of the charge transfer complex Cu:TCNQ. We fabricate a metal-semiconductor-metal device with a single NW and focus ion beam deposited Pt. We observe large photocurrent even at zero bias. The spectral dependence of the photoresponse follows the main absorption at ~ 405 nm which has the primarily responsible for photogenerated carriers. We have quantitatively analyzed the bias dependent photocurrent by a model of two back to back Schottky diodes connected by a series resistance. The observation shows that the large photoresponse of the device primarily occurs due to the reduction of the barrier at the contact regions due to illumination along with the photoconductive contribution. There is also a bias driven reduction of the nanowire resistance that is a unique feature for the material.

In recent years photo-response in single nanowire (NW) of different materials have attracted wide attention for their immense potential for applications in the nanoscale optoelectronic devices. These include nanowires of inorganic semiconducting materials,[1-8] carbon nanotubes,[9] oxide nanowires,[10-13] core shell nanowires,[14,15] polymeric nanowires,[16,17] as well as some organic molecules.[18] These reports showed that it was possible to make photo devices based on a single nanowire, which produced substantial photo-response. In some of the inorganic nanowires like ZnO,[10] the reported photo-response can reach a range of $10^5$ A/W under an applied bias of few volts. One of the reported data on Si nanowire observed generation of substantial photocurrent even at zero applied bias.[8] This gives possibility of making a solar cell as well as unbiased photo detectors. The photo detectors using single nanowires can be made in planar configuration using Schottky type contacts at both ends and they work as metal-semiconductor-metal (MSM) type devices that can be integrated into arrays with peripheral amplifiers built into the same chip.[19]

Very few works has been done on opto-electronic properties and optical switching of CuTCNQ, AgTCNQ and other organometallic compounds' thin film.[22] These organometallic materials had been observed to switch between two stable states when exposed to optical radiation. But, no report has been done on nanowires of such metal-TCNQ charge transfer complex. This paper reports for the first time, a very large photo-response in a single strand of a nanowire of a charge transfer complex Cu:TCNQ (copper tetracyanoquinodimethane) of typical diameter $\leq$ 30nm, using a MSM device configuration with electrodes (separated by $\approx$ 200nm) made from Focused Ion Beam (FIB) deposited Pt. The single NW shows a zero bias current, which with a power density $\sim 2\times10^6$ W/m$^2$ can reach more than 125 nA. At that illumination power density, the enhancement over the dark current at zero bias $>1.25\times10^5$. The effective photo-responsivity at zero bias $\Re(V=0)$, defined as the ratio of current generated to optical power absorbed, was observed around $\approx$ 1-3 A/W and with a bias of 1V applied between two electrodes the observed $\Re$ is well in excess of $10^4$ A/W, which is comparable the highest response seen in such single nanowire devices. The generated photo-current ($I_{Ph}$) shows a maximum at a wavelength close to $\lambda$= 405 nm and the spectral dependence of the $I_{Ph}$ closely mimics the optical absorption in Cu:TCNQ.

Observation of a large photo-response in the charge transfer complex Cu:TCNQ is of importance not only because it is a new addition in the list of single nanowire photo devices that show very large responsivity, but the material itself has a special attribute. Cu:TCNQ is a well known material that shows bias driven resistive state transition from a high resistance state to a low resistance state, when the bias exceeds a critical value and the resistive transition is associated with a memory.[20] This made the material a prime candidate for making MEMISTOR and non-volatile memory.[21] The present report of a very large photo-response, thus adds an opto-electronic element to the existing electronic switching property and raises the possibility of an opto-electronic control to a MEMISTOR device. There are past reports of resistive state switching in Cu:TCNQ films using high power laser illumination[22] and its use as an enabling method for optical writing on this material.[23] In these reports the electric field of the light used $E_{opt}$ is more than the critical field $E_{switch}$ ($\geq 2\times10^6$ V/m) needed to cause the resistive state transition.[22] The phenomena represented in this paper, being controlled mainly by the barriers at contacts and photoconductivity, is distinct from such laser driven resistive state transitions with memory.

RESULTS & DISCUSSION

The Cu:TCNQ nanowires used here were made by vapour phase method by deposition of TCNQ on to a heated Cu-film grown on a glass substrate.[24] (Details are given in Experimental). This leads to formation of Cu:TCNQ nanowires on the Cu-film. The length and diameter of the grown nanowires depend on deposition conditions and time of growth. In the present work, we have grown nanowires with diameters in the range 20-50nm and length of few micrometers. A cross-sectional SEM image of the grown wires on Cu, are shown in inset (a) of Figure 1. The wires were subsequently characterized by a number of techniques that include Micro- Raman Spectroscopy and FTIR spectroscopy to estimate the extent charge transfer and X-Ray diffraction (XRD) to ensure that the Cu:TCNQ nanowires grown are in the more conducting Phase-I (see Experimental). A typical picture of a single strand of a nanowire taken by TEM, on which contacts are made, is shown in Figure 1. The individual Cu:TCNQ nanowires (that grow on the Cu substrate) could be touched with Aluminum coated tip of a conducting probe Atomic Force Microscope (C-AFM). We could measure the bias driven resistive state transition in such wires (length ≈ 1.5 -2.0 µm) occurring at a bias of ≈ 3V corresponding to a field of (1.5-2.0) x $10^6$V/m. (Data in supplementary). This characterization is important to show that even at the highest optical power density used, the electric field of the light used (at λ=405 nm) $E_{opt}$ (~7x$10^4$ V/m) is much less than the applied electric field $E_{switch}$ (≥ 2x$10^6$V/m) needed to switch the resistive state to a low resistance state. (Recent observation[25] of C-AFM based nanocharacterization of Cu:TCNQ nanowires (diameter 20-50nm) with $HfO_2$ barrier/Pt electrode finds switching at bias 5-7 V which is higher than but comparable to the switching bias values seen by us.)

A single nanowire was connected to Cr/Au contact pads by interconnects made of Pt deposited by FIB (FEI -HELIOS 600) using Ga ions (see Experimental) at a voltage of 30 keV and beam current of 90 pA. An image of the sample is shown in the inset (b) of Figure 1. While two inner probes in the single NW device was used for measurements of photo-current, all the 4 electrodes were used to measure the 4-probe resistivity of the nanowire. Our previous analysis[26] of the FIB grown Pt contacts and lines have shown that they are composites of metallic islands consisting of Pt and some Ga (dispersed phase) in a matrix of amorphous carbon. The metallic volume fraction is more than the percolation threshold for metallic conduction making the Pt lines conducting.[26]

In Figure 2 we show the photo-response of the nanowire (NW) device at zero bias, taken at a wavelength λ= 405nm at different incident powers ranging from 98μW to 2.6mW using pulse light. There is no persistent photo-current as the dark current is fully recovered when the illumination is turned off. The power density at the largest power corresponds to $2 \times 10^6$ W/m$^2$. The data shown in Figure 2 at different radiation power densities were taken with a chopper frequency of 130Hz. The zero bias photo current, $I_{Ph}(V=0)$ increases with the optical power density $P_{opt}$ as shown in the inset of Figure 2 and was found to have dependence:

$$I_{Ph}(V=0) \propto P_{opt}^{\gamma} \qquad (1)$$

where the exponent $\gamma \approx 0.3 - 0.4$ for a number of devices tested. In general the power dependence of the photo-current depends on the distribution of trap states around the Fermi level ($E_F$). For systems with traps distributed uniformly with energy around the $E_F$, $\gamma \approx 1$.[27] However, for many photoconductors (in bulk form) $\gamma \approx 1 - 0.5$ showing a non-uniform distribution of trap states around $E_F$.[28] In many nanowires, including measurements done on single nanowires[3,18] the typical exponent was found to be $\gamma \approx 0.3 - 0.5$. It thus appears that in nanowires the traps have a broad distribution of energy around $E_F$. In general the power dependence of the photo-current depends on the recombination kinetics and the distribution of trap states around the Fermi level ($E_F$). The large carrier generation rate in a confined volume is expected to increase the non-geminate recombination cross-section and reduce $\gamma$.

For the device under test (DUT) [nanowire of diameter $d_{NW} = 30$ nm, with a length ($l$) between electrodes $\approx 200$ nm], assuming the NW absorbs all around and over the whole length $l$ between the electrodes, an optical power density of $P_{opt} = 1$ W/m² corresponds to $1.9 \times 10^{-14}$ W absorbed by the sample between two electrodes. Since the absolute value of the absorbance is not known this is the maximum power absorbed by the sample. The photo-response at zero bias $I_{Ph}(V=0)$ were measured at different chopping frequencies at $\lambda = 405$ nm (data in supplementary). The response increases with the frequency at low frequency and then saturates beyond 2 kHz. For a chopper frequency of 2 kHz and $P_{opt} = 2 \times 10^6$ Watt/m² the responsivity at zero bias $\Re(V=0) \equiv \frac{I_{ph}(V=0)}{\tilde{P}_{sample}} \approx$ 3 A/W. This compares very well with GaN NW[6] [$\Re(V=0) \approx 0.1$ A/W] that shows one of the highest recorded zero bias responsivity and much higher than that seen in Si single NW[8] [$\Re(V=0) \approx 0.6$ mA/W].

Due to the sub-linear dependence of $I_{Ph}(V=0)$ on optical power density $P_{opt}$ (Equation (1)), $\Re(V=0)$ decreases as $P_{opt}$ increases. For continuous illumination at $\lambda = 405$ nm, at $P_{opt} \approx 6.6 \times 10^6$ W/m², $\Re(V=0) = 1.6$ A/W for the same DUT. However, application of bias enhances the responsivity $\Re$ significantly and we could reach a value of $8 \times 10^4$ ($4 \times 10^3$) A/W for an applied bias of 1 (-1) V. This is an enormous current response from a single nanowire and compares very well with some of the highest reported $\Re$ under comparable bias (e.g, ZnO) [10] that show at 1V $\Re \approx 10^5$ A/W. The $I-V$ data on the device taken in dark and under continuous illumination ($\tilde{P}_{sample} \leq 3.0 \times 10^{-10}$ Watt, $\lambda$ = 405nm) are shown in Figure 3.

The wavelength dependence of the photo-response $I_{Ph}(\lambda)$ is shown in Figure 4 in the wavelength range 300-900 nm. The data are taken with a small bias of V= 0.5V. The data shows a peak around $\lambda \approx 405$ nm. In the inset of Figure 4, we show the absorption curve. The absorption data were taken in two methods. One method is by dispersing the nanowires in ethanol by ultrasonication, and other method is by dispersing the nanowires on a polished quartz plate and then allow the dispersing medium to evaporate. In both cases the normalized absorption was found to be similar with a maximum at around the same $\lambda$ ($\approx$405 nm) and broad and shallow maxima at longer wavelengths around 700 nm. The absorption data recorded by us matches well with the reported absorption spectra.[23,29] Comparison with the absorption and the optical response $I_{Ph}(\lambda)$ curves show that the optical response is indeed caused by the absorption of the light in the Cu:TCNQ nanowire predominantly at the peak absorption range, while at longer wavelength the response is at least an order less. The observed photo-response is thus due to creation of carriers (electron–hole pairs) at the absorption peak. It is generally believed that on illumination the light is absorbed by the TCNQ moiety which leads to creation of neutral TCNQ and the absorption peak in Cu:TCNQ arises mainly from neutral (anion radical) TCNQ (TCNQ$^{-1}$).[29] If that indeed is the case, there is creation of electron from the TCNQ anion radical.

The $I-V$ data in dark as well as under continuous illumination at 405nm are shown in Figure 3. The dark as well as illuminated $I-V$ data have been analyzed in the frame work of a metal-semiconductor-metal (MSM) structure where the FIB deposited Platinum electrodes act as the metal electrodes and the Cu:TCNQ is the semiconductor. The current through the device has a zero bias component (under illumination) as well as a bias dependent current $I(V)$. The bias dependent current through the device is fitted to the model of two back to back Schottky diodes connected by a series resistance R, which mainly represents the resistance due to the portion of the NW between the electrodes. The equation for fit used is:[30]

$$I(V) = I_0 \exp\left(\frac{qV'}{\eta kT} - 1\right) \frac{\exp\left(\frac{-q(\phi_1 + \phi_2)}{kT}\right)}{\exp\left(\frac{-q\phi_2}{kT}\right) + \exp\left(\frac{-q\phi_1}{kT}\right)\exp\left(\frac{qV'}{\eta kT}\right)} \qquad (2)$$

where, $V' = V - IR$, $R$ being the series resistance, and $\phi_1$ and $\phi_2$ are the barrier heights associated with two contacts (M's). In the equation above $\phi_1$ refers to terminal with V +ve. $I_0$ arises from thermoinic emission and does not change on illumination. (The bias dependent $I(V)$ is added to experimentally observed zero bias photocurrent $I_{Ph}(V = 0)$ to obtain the total device current $I$). Ideally if the two M contacts in the MSM structure are identical, the $I-V$ curves will be symmetric. However, unequal barrier heights $\phi_1 \neq \phi_2$ at the two contacts will lead to asymmetric $I-V$ curves. Figure 5 shows the schematic device structure and the band diagram for dark and illuminated conditions. From the fit of the experimental $I-V$ data using equation 2, we obtain $\phi_1 \approx 0.27$ eV and $\phi_2 \approx 0.45$ eV in the dark with $R = 1$ kΩ (solid line through the data in Figure 3 shows the fit to the $I-V$ curve). The series resistance $R$ arises mainly because of the portion of the nanowire between electrodes and corresponds to a resistivity $\rho \approx 0.36 \times 10^{-5}$ Ohm-m. This about a factor of 8 less than the value of $\rho$ measured by 4-probe method using a low bias <0.2V. This, however is comparable to the value of $\rho$ measured by the same method at a bias > 0.5V. This shows that the applied bias > 0.5V leads to reduction in the resistivity. This is likely can occur because beyond a bias of 0.5 V, the applied field $2.5 \times 10^6$ V/m, which is higher than the $E_{switch}$ ( ~$2 \times 10^6$ V/m). Thus in the substantial range of bias in which the $I-V$ curve, has been taken, there is a voltage driven resistance reduction though not a switching with memory. [Switching with memory generally occurs when one of the electrode is an oxide.[21]]

On illumination at 405nm, the barrier heights change to lower value, $\phi_1 \approx 0.18$ eV and $\phi_2 \approx 0.34$ eV and $R = 0.6$ k$\Omega$. Both the barrier heights are thus reduced by nearly the same amount of 0.1 eV on illumination and the series resistance also is suppressed by nearly a factor of 2. The suppression of the series resistance $R$ is due to the photoconductivity in the bulk of the nanowire between the two contacts.

The high responsivity seen in Schottky type photodiodes (or in MSM type devices) depend on the light induced barrier reduction in the contact region, a concept that was known for long.[31] It had been shown that in MSM type devices made on nanowires, one can obtain lowering of barrier, if the illumination can create enough carriers in the contact region.[8,32,33] Control of contact barrier by electrical gate controlled charge injection has been recently utilized in Si NW devices.[34] Lowering of interfacial barriers on illumination at the metal-semiconductor junction in polymer field effect transistors, have been investigated using optical excitation directed in the electrode region without optically perturbing the channel.[35] Even charge generated in piezo-electric ZnO NW was found to lower the barrier.[11] The Pt electrodes used by us have thickness (~100nm) that is lower than the skin depth (~205 nm) of the electrode at the wavelength 405 nm. As a result, there will be light induced carrier generation under the contact region leading to reduction in the barrier at the contacts. The analysis of the $I-V$ data thus leads to the conclusion that lowering of barrier at the two M-contacts under illumination is an important, if not the dominant, mechanism, for large photo-response.

The photocurrent has contribution of illumination induced reduction of the barrier at the junction, as well as photoconductive reduction of the series resistance of the wire. To test the relevant contributions we calculated the photocurrent (using eqn. 2) when barrier heights under illumination are used and the series resistance is kept at the value under dark. This is shown by dashed-dot lines in Figure 3. The reduction of the barrier height leads to increase of the current; however, there is effect also from the resistance R. When the barrier height is not changed but only the R under illumination is used for calculation (effect of photoconductivity alone), the calculated photocurrent does not show any appreacibale change (dashed line in Figure 3). This shows that the lowering of the barrier at the contact under illumination is the primary reason for the photo response, which in turn accentuates the effect of the photoconductivity.

The asymmetry in the $I-V$ data can be explained by a small difference in the work functions at the two electrodes which can be due to small variations in the exact composition of carbon and Pt/Ga in the FIB deposited Pt. This asymmetry can also occur due to difference in size of the two electrodes.[35] It has been shown that the different sizes of electrodes in an MSM junction can lead to asymmetry because of the different current densities (for the same device current) that may affect the band bending and the depletion layer width in the junction region leading to asymmetry.

The diameter of the NW being ~30nm and the photocurrent $I_{Ph}(V)$ observed under a bias of 1V was in excess of 0.01 A. This corresponds to a very large current density $J_{Ph} \approx 1.4 \times 10^{13}$ A/m$^2$. The photocurrent density is much larger than that observed in the electrically switched state in the single nanowire, which is typically 100 nA, corresponding to $J_{Ph} \approx 1.4 \times 10^8$ A/m$^2$ (see supplementary). It is noteworthy that the material can indeed carry such a large current density (in excess of $10^{13}$ A/m$^2$) without getting damaged.

The response that is seen here is not due to thermal effects as may arise due to heating of the substrate with the radiation use. Our estimate shows that the rise in temperature is not more than few degrees and the response observed cannot rise from thermal effects.

In a preliminary experiment we find, an adequate photocurrent can be observed when an array of such nanowires (made by lateral growth between a pair of Cu electrodes with separation ~1µm) are exposed to similar illumination. However, the magnitude of the current is smaller than that observed in the present single nanowire device. Thus photo-generation is a generic property of the Cu:TCNQ nanowires, irrespective of the nature of electrodes, although the value of the carrier generated depends on the nature of electrodes that control the barrier at the contacts.

There is an unambiguous observation of zero bias photo current in this material. One of the mechanisms proposed for transport of photo-generated carriers in MSM junction is the tunnelling of photoexcited carriers by tunnelling through the barrier at the junction.[37] This is prevalent when the photon energy $h\nu$ is much greater than the barrier height $\phi$, which is adequately satisfied in our case. Existence of such a tunnelling of the photoelectron at the junction can be a source of the photocurrent at zero bias.

CONCLUSSION: In summary, we observed very large photoresponse in a single NW of the charge transfer complex Cu:TCNQ. This is the first photoresponse in NW made of such a material. The single NW was made as a metal-semiconductor-metal device using FIB deposited Pt. The observed photo current has a spectral dependence that strongly follows the main absorption (close to 405nm) showing the primary role of the photo-generated carriers. There is an observed current also in zero bias and the responsivity increases from ≈ 3 A/W at zero bias to excess of $10^4$ A/W at a modest bias of 1V. The observed photoresponse with bias is comparable to the largest seen in some other single NW photodectors. The photocurrent under bias primarily occurs due to the reduction of the barrier at the contact regions due to illumination along with the photoconductive contribution. It is found that for photo response under bias, there is also a bias driven reduction of the NW resistance, a feature that is unique for the material. In addition, to a very large optical response, the observation also adds an important opto-electronic functionality to the charge transfer complex that is know for resistive state switching and for non-volatile RAM applications.

METHODS

*Growth:* The Cu:TCNQ nanowires were grown by vapour phase deposition method[24] on Cu film (kept at ~130$^o$C) using TCNQ powder as a source. The diameters of the nanowires grown by this method ranges from 20 to 60 nm. (See Supplementary information for images). The characterization by XRD shows that the nanowires are grown in Phase-I which has a tetragonal unit cell[38] (XRD data in supplementary). The charge transfer(Z) was measured by shift of C-CN strech vibration at 1376 cm$^{-1}$ as measured by Raman Scattering and was found to be Z≈0.97. For measuring Z the calibration of shift in C-CN strech band with Z has been used.[39] In a separate test, to check that these wires grown have the proper electrical characteristics that they can show resistive switching, we measured the resistance of an individual nanowire (grown on Cu) using a conducting probe Atomic Force Microscope (C-AFM) and found they show resistive switching at a bias of ≈ 3.5V when the probe has a pre-evaporated Al/Al$_2$O$_3$ coating. (see Supplementary information). The observed swicthing characteristics are similar to that measured with CAFM in stacks with Cu:TCNQ nanowires with HfO$_2$ barrier.[25]

*Fabrication of the device:* For attaching leads for the photo response measurements, the nanowires were dispersed on SiO$_2$ (300 nm thick)/Si (insulating) substrate containing pre-fabricated Cr/Au electrodes which were used as contact pads. Dispersing medium used was ethanol. The operation of attching leads to individual nanowires were done in a dual beam machine FEI HELIOS 600. The wires were imaged during fabrication, using FE-SEM (Field-emission scanning electron microscope). After selection of a single nanowire 4 leads were then connected using Pt electrodes deposited by FIB at a voltage of 30 keV and deposition current of 90 pA . The nature of the FIB deposited Pt was investigated before and was found to have a room temperature resistivity below 10 $\mu\Omega$-m which is more or less temperature independent.

*Sources of illumination:* The photo-response at a single wavelength (405nm) was measured using a diode laser and a microscope that allows simultaneous observation on the placement of the light spot. The power dependence was measured by changing the power using a set of neutral density filter. The wavelength dependence of the photo-current as well as the I-V curve under illumination were also measured using a Helium lamp (450W) and a monochromator. The I-V curves were recorded using a source meter.

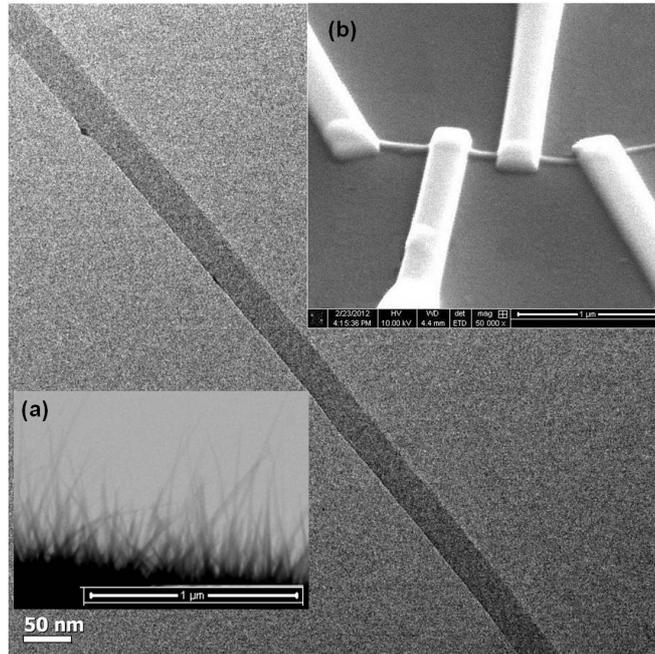

**Figure 1.** TEM image of a typical Cu:TCNQ nanowire of diameter 30 nm. Inset (a) shows an FE-SEM image of vertical growth of the nanowire on Cu film. The inset (b) shows an FE-SEM image of a 30 nm single Cu:TCNQ nanowire device connected with four FIB deposited Pt leads.

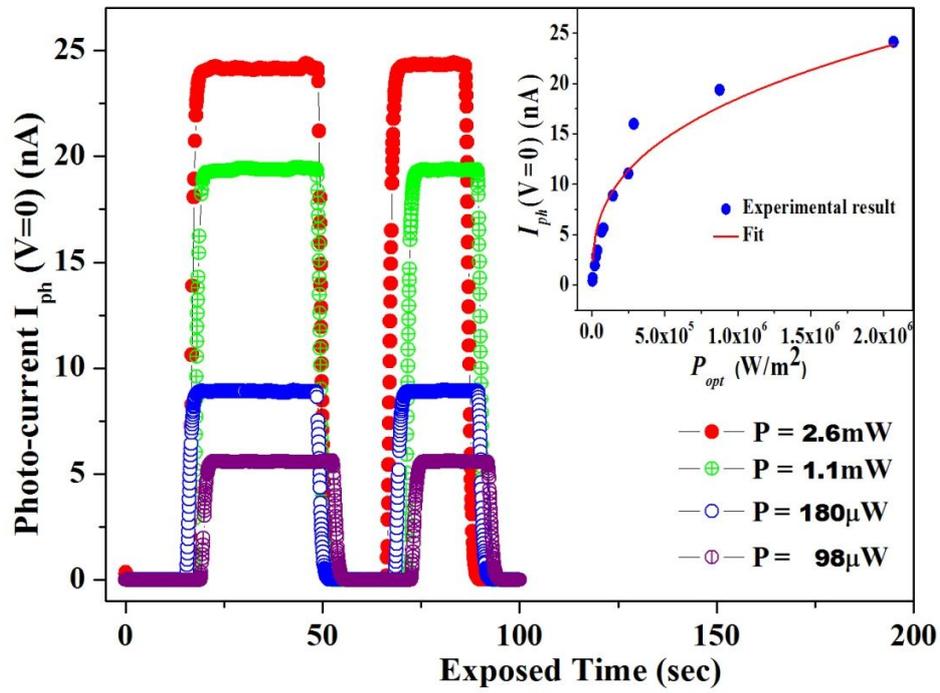

**Figure 2.** Reversible zero bias photoresponse $I_{ph}$ (V=0) of a 30nm Cu:TCNQ nanowire under an illumination at wavelength 405nm that was turned on and off. Inset shows zero bias photocurrent at different optical power density $P_{opt}$ with an excitation wavelength of 405nm.

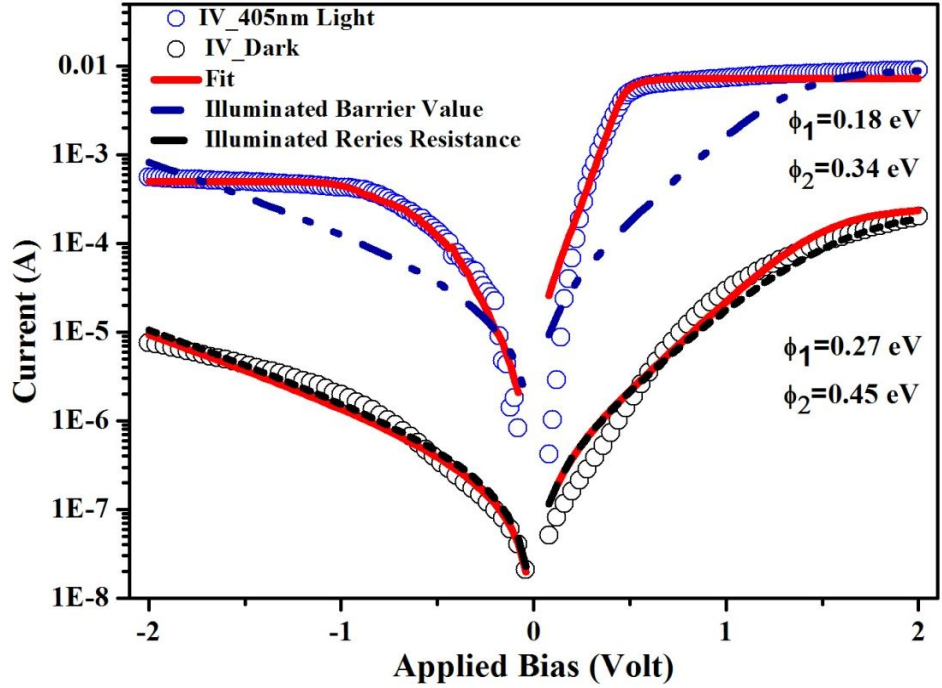

**Figure 3**. *I-V* curves in dark (black circle) and under illumination (blue circle) of a single Cu:TCNQ nanowire. Wavelength of illumination is 405 nm. Solid red curve is fit to the MSM model. The barrier heights $\Phi_1$ and $\Phi_2$ obtained from the fit are $\Phi_1= 0.27eV$, $\Phi_2=0.45eV$ in dark and $\Phi_1= 0.18eV$, $\Phi_2=0.34eV$ under illumination. The dashed-dotted (a) curve and dashed (b) curve have been calculated with (i) dark series resistance value but illuminated barrier values and with (ii) illuminated series resistance value but dark barrier values respectively.

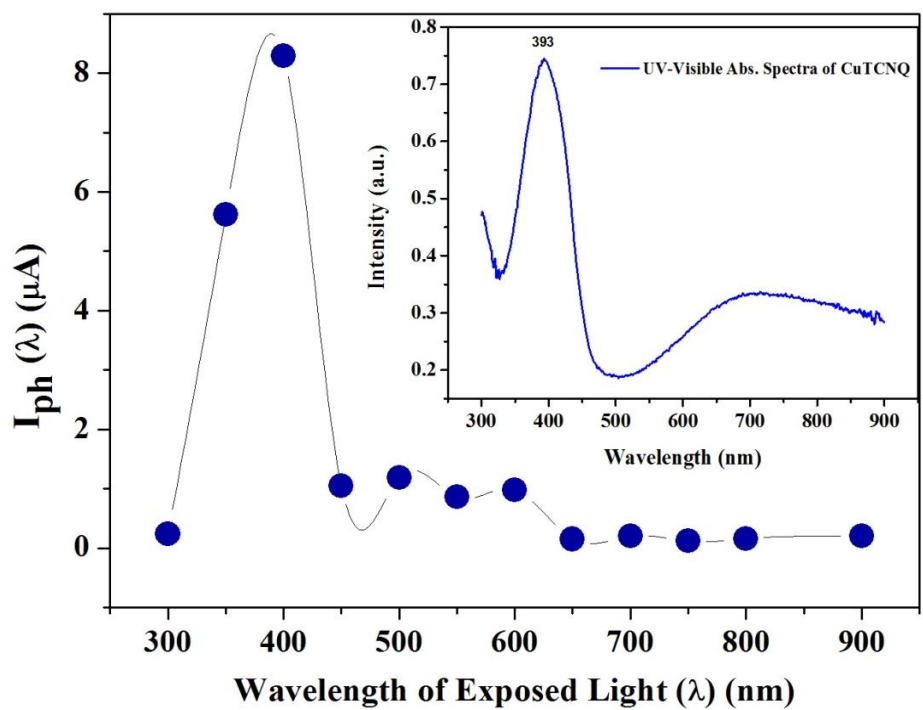

**Figure 4**. Photocurrent ($I_{ph}$) spectral response of the single nanowire device kept at a constant bias of *0.5 V*. Photo-current is maximum at λ ~ 400nm. The inset shows the absorption spectra of Cu:TCNQ nanowires.

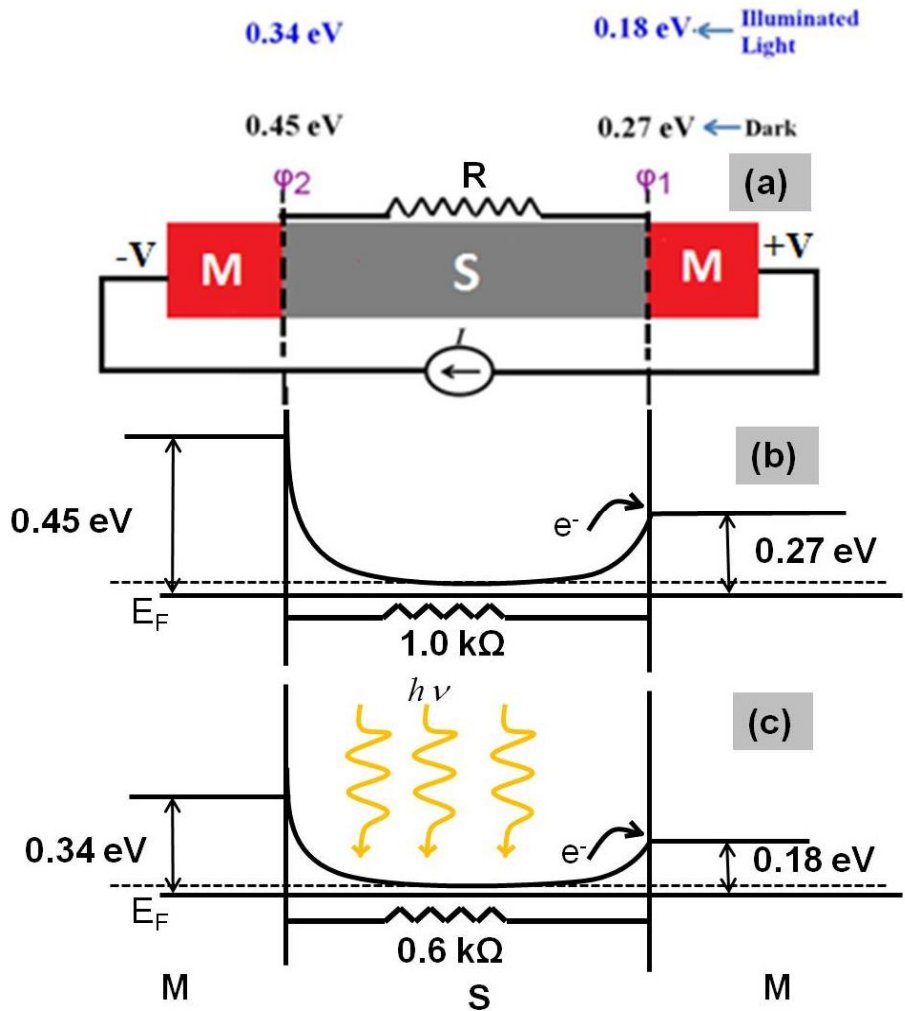

**Figure 5.** (a) Schematic diagram of MSM structure. Band diagram of the structure (b) before illumination, the barrier heights in forward and reverse bias are $\Phi_1$=0.27 eV and $\Phi_2$=0.45 eV respectively.(c) As light is illuminated barrier height reduces to $\Phi_1$=0.18 eV and $\Phi_2$=0.34 eV. The asymmetry in the I-V curve arises from the different barrier heights at the two contacts.

## ASSOCIATED CONTENT

**Supporting Information:** Electrical resistive switching of single nanowire of Cu:TCNQ, Dependence of chopper frequency on photo-current of Single Cu:TCNQ nanowire, Structural characterization (SEM , X-RD) This material is available free of charge via the Internet at http://pubs.acs.org

## ACKNOWLEDGMENT

The work is supported by the Department of Science, Government of India as a Theme Unit of Excellence in Nanodevice Technology under Nanomission. Partial support for project personnel was from the project UNANST-II of the Department of Science and Technology, Government of India. Author R. Basori wants to thank Dr. Sudeshna Samanta and Mr. Shahnewaz Mondal for useful discussions on C++ programming.

REFERENCES

1. Wang J.; Gudiksen M. S.; Duan X.; Cui Y., Lieber C. M. Highly Polarized Photoluminescence and Photodetection from Single Indium Phosphide Nanowires, *Science* **2001**, *293*, 1455-1457.

2. Chen R-S.; Yang T-H.; Chen H.; Chen L-C.; Chen K-H.; Yang Y-J.; Su C-H.; and Lin C-R. High-gain photoconductivity in semiconducting InN nanowires, *Appl. Phys. Lett* . **2009**, *95*, 162112-3.

3. Kun, S-C.; Xing, W.; van der Veer, W- E.; Yang, F.; Donavan, K-C.; Cheng, M.; Hemminger, J-C.; and Penner,R-M. Tunable Photoconduction Sensitivity and Bandwidth for Lithographically Patterned Nanocrystalline Cadmium Selenide Nanowires *ACS Nano* **2011,** *5*, 7627-7639.

4. Huang, H. M.; Chen, R. S.; Chen, H. Y.; Liu, T. W.; Kuo, C. C. Photoconductivity in single AlN nanowires by subband gap excitation,*Appl. Phys. Lett*. **2010,** *96*, 062104-3.

5. Han, S.; Jin, W.; Zhang, D. H.; Tang, T.; Li, C.; Liu, X. L.; Liu, Z. Q.; Lei, B.; Zhou, C. W. Photoconduction studies on GaN nanowire transistors under UV and polarized UV illumination. *Chem. Phys. Lett.* **2004**, *389*, 176-180.

6. Chen, R-S.; Wang, S-W.; Lan, Z-H.; Tsai, J. T-H.; Wu, C-T.; Chen, L-C.; Chen, K-H.; Huang, Y-S.; and Chen, C-C. On-Chip Fabrication of Well-Aligned and Contact-Barrier-Free GaN Nanobridge Devices with Ultrahigh Photocurrent Responsivity. *Small* **2008**, *4*, 925-929.

7. Calarco, R.; Marso, M.; Richter, T.; Aykanat, A. I.; Meijers,R.; Hart, A.v.d.; Stoica, T.; and Luth , H. Size-dependent Photoconductivity in MBE-Grown GaN-Nanowires. *Nano Lett.,* **2005**, *5*, 981-984.

8. Ahn, Y.; Dunning, J.; Park,J. Scanning Photocurrent Imaging and Electronic Band Studies in Silicon Nanowire Field Effect Transistors. *Nano Lett*. **2005**, *5*, 1367-1370.

**9.** Freitag, M.; Martin, Y.; Misewich, J. A.; Martel, R.; Avouris, P. Photoconductivity of Single Carbon Nanotubes. *Nano Lett.* **2003**, *3*, 1067-1071.

**10.** Kind, H.; Yan, H.; Messer, B.; Law, M.; Yang, P. Nanowire Ultraviolet Photodetectors and Optical Switches. *Adv. Mater.* **2002**, *14*, 158-160.

**11.** Liu, Y.; Yang, Q.; Zhang, Y.; Yang, Z.; and Wang, Z. L. Nanowire Piezo-phototronic Photodetector: Theory and Experimental Design. *Adv. Mater.* **2012**, *24*, 1410-1417.

**12.** Chen, M-W.; Retamal, J. R. D.; Chen, C-Y.; He, J-H. Photocarrier Relaxation Behavior of a Single ZnO Nanowire UV Photodetector: Effect of Surface Band Bending. *IEEE ELEC. DEV. LETT.* **2012**, *33*, 411-413.

**13.** Huang, K.; Zhang, Q.; Yang, F.; He, D. Ultraviolet Photoconductance of a Single Hexagonal WO3Nanowire. *Nano Res.* **2010**, *3*, 281-287.

**14.** Afsal, M.; Wang, C-Y.; Chu, L-W.; Ouyang, H.; Chen, L-J. Highly sensitive metal–insulator–semiconductor UV photodetectors based on ZnO/SiO2 core–shell nanowires. *Journal of Mater.Chem* **2012,** *22*, 8420-8425.

**15.** Li, Z.; Rochford, C.; Baca, F. J.; Liu, J.; Li, J.; Wu, J. Investigation into Photoconductivity in Single CNF/TiO2-Dye Core–Shell Nanowire Devices *Nanoscale Res. Lett.* **2010**, *5*, 1480-1486.

**16.** O'Brien, G. A.; Quinn, A. J.; Tanner, D. A.; Redmond, G. A Single Polymer Nanowire Photodetector. *Adv. Mater.* **2006**, *18*, 2379-2383.

**1**7. Xie,X. N.; Xie, Y.; Gao, X.; Sow, C. H.; Wee, A.T. S. Metallic Nanoparticle Network for Photocurrent Generation and Photodetection. *Adv. Mater.* **2009**, *21*, 3016-3021.

**18.** Zhang, X.; Jie, J.; Zhang, W.; Zhang, C.; Luo, L.; He, Z.; Zhang, X.; Zhang, W.; Lee, C.; Lee, S. Photoconductivity of a Single Small-Molecule Organic Nanowire. *Adv. Mater.* **2008**, *20*, 2427-2432.

**19.** Fan, Z.; Ho, J. C.; Jacobson, Z. A.; Razavi, H.; Javey, A. Large-scale, heterogeneous integration of nanowire arrays for image sensor circuitry. *Proc. Nat.Acad.Sci.* **2008**, *105*, 11066-11070.

**20.** Potember, R.S.; Poehler, T.O.; Cowan, D.O. Electrical switching and memory phenomena in CuTCNQ thin films. *Appl. Phys. Lett.* **1979**, *34*, 405-407.

**21.** Müller, R.; Jonge, S. D.; Myny, K.; Wouters, D. J.; Genoe, J.; Heremans, P. Organic CuTCNQ integrated in complementary metal oxide semiconductor copper back end-of-line for nonvolatile memories *Appl. Phys. Lett.* **2006**, *89*, 223501-3**.** and Mueller, R.; Billen, J.; Katzenmeyer, A.; Goux, L.; Wouters, D. J.; Genoe, J.; Heremans, P. Resistive Electrical Switching of $Cu^+$ and $Ag^+$ based Metal-Organic Charge Transfer Complexes. *Mater. Res. Soc. Symp. Proc.* **2008,** *1071*, F06-04.

**22.** Potember, R.S.; Poehler, T.O.; Benson, R.C. Optical switching in semiconductor organic thin films. *Appl. Phys. Lett.* **1982**, *41*, 548-550.

**23.** Huang, W-Q.; Yi, Q- W.; Gu, D-H.; Gan, F-X. Green-Light Static Rewritable Optical Storage Properties of a Novel CuTCNQ Derivative Thin Film. *Chinese Phys. Lett.* **2003**, *20*, 2178-2181.

**24.** Oyamada, T.; Tanaka, H.; Matsushige, K.; Sasabe H.; Adachi, C. Switching effect in Cu:TCNQ charge transfer-complex thin films by vacuum codeposition *Appl. Phys. Lett.* **2003**, *83*, 1252-1254.

**25.** Deleruyelle, D.; Muller, C.; Amouroux, J.; Müller, R. Electrical nanocharacterization of copper tetracyanoquinodimethane layers dedicated to resistive random access memories. *Appl. Phys. Lett.* **2010**, *96*, 263504-3.

**26.** Chakravorty, M.; Das, K.; Raychaudhuri, A.K.; Naik, J.P.; Prewett, P.D. Temperature dependent resistivity of platinum–carbon composite nanowires grown by focused ion beam on SiO2/Si substrate. *Microelectronic Engg.* **2011**, *88*, 3360-3364.

**27.** A.Rose, in *Concepts in Photoconductivity and Allied Problems*, Vol 19 Interscience Publishers, New York **1963**.

**28.** R.H.Bube, in *Photoelectronic properties of Semiconductors*, Cambridge University Press, Cambridge **1992**.

**29.** Liu, S-G.; Liu, Y-Q.; Wu, P-J.; Zhu, D-B. Multifaceted Study of CuTCNQ Thin-Film Materials. Fabrication, Morphology, and Spectral and Electrical Switching Properties. *Chem. Mater.* **1996,** *8,* 2779-2787.

**30.** Norde, H. A modified forward IV plot for Schottky diodes with high series resistance *J.Appl.Phys.* **1979**, *50*, 5052-5053.

**31.** Schneider, M.V. Bell System Tech Journal **1966**, *45*, 161 and Mehta R.R.; Sharma, B.S. Photoconductive gain greater than unity in CdSe films with Schottky barriers at the contacts *J. Appl. Phys.* **1973**, *44*, 325-328.

**32.** Wei, T-Y.; Huang, C-T.; Hansen, B. J.; Lin, Y-F.; Chen, L-J. Large enhancement in photon detection sensitivity via Schottky-gated CdS nanowire nanosensors. *Appl. Phys. Lett*. **2010**, *96*, 013508-3.

**33.** Gu, Y.; Kwak, E-S.; Lensch, J-L.; Allen, J-E.; Odom, T-W. Near-field scanning photocurrent microscopy of a nanowire photodetector. *Appl. Phys. Lett*. **2005**, *87*, 043111-3.

**34.** Mongillo, M.; Spathis, P.; Katsaros, G.; Gentile P.; Franceschi, S. D. Multifunctional Devices and Logic Gates With Undoped Silicon Nanowires. *Nano Lett*. **2012**, *12*, 3074-3079.

35. Rao M.; Narayan, K. S. Evaluation of electrode-semiconductor barrier in transparent top-contact polymer field effect transistors. *Appl. Phys. Lett.* **2008**, *92*, 223308-3.

36. Okyay, A. K.; Chui, C. O.; Saraswat, K. C. Leakage suppression by asymmetric area electrodes in metalsemiconductor-metal photodetectors. *Appl. Phys. Lett.* **2006,** *88*, 063506-2.

37. Diesing, D.; Merschdorf, M.; Thon, A.; Pfeiffer, W. Identification of multiphoton induced photocurrents in metal–insulator–metal junctions *Appl. Phys.B, Laser and Optics.* **2004**,*78*, 443-446.

38. Robert, A.; Zhao, H.; Ouyang, X.; Grandinetti, G.; Cowen, J.; Dunbar, K.R. New Insight into the Nature of Cu(TCNQ): Solution Routes to Two Distinct Polymorphs and Their Relationship to Crystalline Films That Display Bistable Switching Behavior *Inorg. Chem.*, **1999**, *38*, 144-156.

39. Kamitsos, E. I.; Risen, W. M. Raman studies in CuTCNQ: Resonance Raman spectral observations and calculations for TCNQ ion radicals *J. Chem. Phys.* **1983,** *79*, 5808-5819.